\documentclass[12pt,preprint]{aastex}
\usepackage{psfig}

\newcommand\pp     {$\pm$}

\newcommand\Lunit   {erg s$^{-1}$}
\newcommand\funit   {erg cm$^{-2}$ s$^{-1}$}

\righthead{KS 1731--260 in quiescence}
\slugcomment{Accepted for publication in ApJ Letters: September 13, 2001}

\begin{document}

\title{A {\itshape Chandra} observation of the long-duration X-ray
transient KS 1731--260 in quiescence: too cold a neutron star?}

\author{Rudy Wijnands\altaffilmark{1,4}, Jon
M. Miller\altaffilmark{1}, Craig Markwardt\altaffilmark{2}, Walter
H. G.  Lewin\altaffilmark{1}, Michiel van der Klis\altaffilmark{3}}

\altaffiltext{1}{Center for Space Research, Massachusetts Institute of
Technology, 77 Massachusetts Avenue, Cambridge, MA 02139-4307, USA;
rudy@space.mit.edu}

\altaffiltext{2}{NASA/Goddard Space Flight Center, Code 662,
Greenbelt, MD 20771}

\altaffiltext{3}{Astronomical Institute ``Anton Pannekoek'',
       University of Amsterdam, Kruislaan 403, NL-1098 SJ Amsterdam,
       The Netherlands}

\altaffiltext{4}{Chandra Fellow}

\begin{abstract}

After more than a decade of actively accreting at about a tenth of the
Eddington critical mass accretion rate, the neutron-star X-ray
transient KS 1731--260 returned to quiescence in early 2001. We
present a {\it Chandra}/ACIS-S observation taken several months after
this transition. We detected the source at an unabsorbed flux of $\sim
2 \times 10^{-13}$ \funit~(0.5--10 keV).  For a distance of 7 kpc,
this results in a 0.5--10 keV luminosity of $\sim 1 \times 10^{33}$
\Lunit~ and a bolometric luminosity approximately twice that.  This
quiescent luminosity is very similar to that of the other quiescent
neutron star systems.  However, if this luminosity is due to the
cooling of the neutron star, this low luminosity may indicate that the
source spends at least several hundreds of years in quiescence in
between outbursts for the neutron star to cool. If true, then it might
be the first such X-ray transient to be identified and a class of
hundreds of similar systems may be present in the Galaxy.
Alternatively, enhanced neutrino cooling could occur in the core of
the neutron star which would cool the star more rapidly. However, in
that case the neutron star in KS 1731--260 would be more massive than
those in the prototypical neutron star transients (e.g., Aql X-1 or 4U
1608--52).

\end{abstract}


\section{Introduction \label{section:intro}}

The X-ray transients are a special sub-group of low-mass X-ray
binaries (LMXBs).  These systems are usually very dim, with X-ray
luminosities of $10^{30-34}$ \Lunit, but they exhibit sporadic
outbursts during which the luminosity increases to $10^{36-39}$
\Lunit. The exact mechanisms for the quiescent X-ray emission are
still uncertain (e.g., Menou et al. 1999; Campana \& Stella 2000;
Bildsten \& Rutledge 2001).  The most promising and testable model for
the quiescent properties of neutron star transients is the one which
assumes that the X-rays below a few keV\footnote{In the spectra of
several systems, a power law component above a few keV is present
(e.g., Asai et al. 1996, 1998) which is not understood but which may
be due to residual accretion (see Campana \& Stella 2000 for a
discussion).  Here, we will not discuss this component.}  are due to
cooling of the neutron star after the accretion has stopped (van
Paradijs et al. 1987; Verbunt et al. 1994; Asai et al. 1996; Brown,
Bildsten, \& Rutledge 1998). The main argument against this
interpretation used to be that the temperature obtained from
black-body fits to the spectral data resulted in very small ($<$3 km)
emitting radii, significantly smaller than the theoretical predictions
for neutron star radii. However, recently it has been put forward that
black-body fits to the spectra overestimated the effective
temperatures and thus underestimated the emitting areas (Brown et
al. 1998). The spectra are better and more realistically described by
thermal emission from a pure hydrogen atmosphere.  By using data of
quiescent neutron star systems, Rutledge et al. (1999, 2000)
demonstrated that their spectra could indeed be fitted with such
hydrogen atmosphere models and emission radii of $\sim$10 km were
obtained.

If the quiescent emission below a few keV is indeed due to the cooling
of the neutron star, then the actual luminosity is predicted to depend
on the time-averaged accretion rate (Brown et al. 1998).  Using their
cooling model, a distance independent prediction of the expected
quiescent flux can be derived of $ F_{\rm q} \approx \langle F
\rangle/130 $ (see also Rutledge et al. 2001b) with $\langle F
\rangle$ the time averaged flux of the source. The latter can be
rewritten as $\langle F \rangle = t_{\rm o} \langle F_{\rm o} \rangle
/ (t_{\rm o} + t_{\rm q})$ resulting in $F_{\rm q} \approx {t_{\rm o}
\over t_{\rm o} + t_{\rm q}} \times {\langle F_{\rm o} \rangle \over
130}$, with $\langle F_{\rm o} \rangle$ the average flux during
outburst, $t_{\rm o}$ the average time the source is in outburst, and
$t_{\rm q}$ the average time the source is in quiescence. This model
indeed appears to reproduce the observed quiescent luminosities (e.g.,
Rutledge et al. 1999, 2000). Furthermore, Brown et al. (1998) also
predicted the very low luminosity of SAX J1808.4--3658 in quiescence
(Dotani, Asai, \& Wijnands 2000; Wijnands et al. 2001a) due to its
very low time averaged accretion rate.  Despite these successes, more
quiescent systems should be observed to test and refine the model
further.

A class of sources not yet compared in this model is that of the
long-duration transients (e.g., MXB 1659--298; EXO 0748--676; KS
1731--260 [see below]). These are systems which suddenly, like
ordinary transients, become X-ray bright but in contrast to the
ordinary, short-duration ones, do not disappear after a few weeks to
months, but remain active for several years to decades. If the only
difference between those systems and the ordinary ones is the duration
$t_{\rm o}$ of the outburst episodes, and everything else is similar,
in particular if the time $t_{\rm q}$ the sources spend in quiescence
is similar to that of normal transients, then the neutron stars of
such long-duration transients are predicted to be heated to
considerably higher temperatures than those in the ordinary ones. Up
to recently, this prediction could not be tested because none of the
known long-duration neutron star transients which are currently active
had turned off. In this {\it Letter}, we discuss the first results
obtained of such a system, KS 1731--260, which turned off in early
2001 after having accreted for over a decade.

\begin{figure}
\begin{center}
\begin{tabular}{c}
\psfig{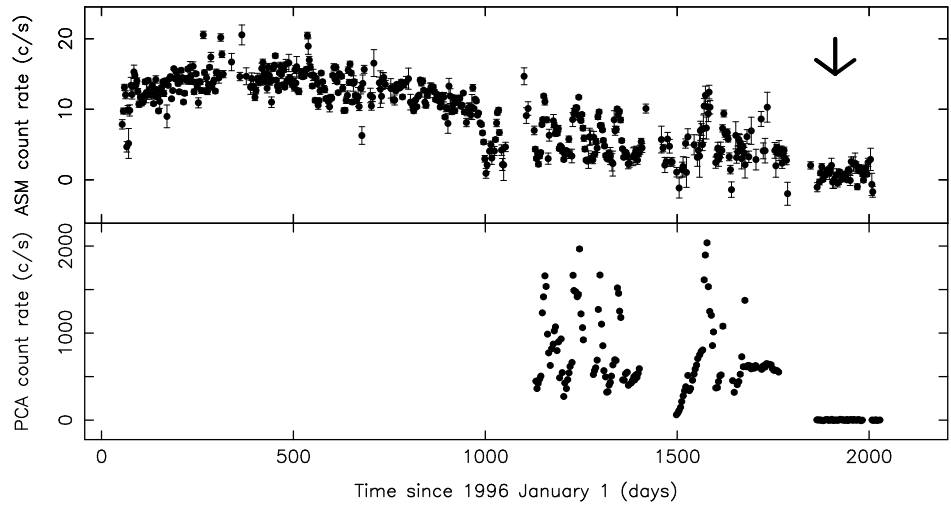}
\end{tabular}
\figcaption{The {\it RXTE}/ASM 1.5--12 keV (top) and the {\it
RXTE}/PCA (bottom) 2--60 keV count rate curves of KS 1731--260. The
day the {\it Chandra} observation was performed is marked with an
arrow.
\label{fig:lc} }
\end{center}
\end{figure}

KS 1731--260 was discovered in Aug. 1989 using the {\it Mir}/Kvant
instrument and was also found to be active in Oct. 1988 (Sunyaev
1989). The compact object in this transient is a neutron star as it
exhibits type-I X-ray bursts (Sunyaev 1989; Sunyaev et
al. 1990). Since its initial discovery, its has been observed as a
persistent source with {\it Ginga}, {\it Sigma}, {\it ROSAT}, {\it
RXTE}, and {\it ASCA} (Yamauchi \& Koyama 1990; Barret et al. 1992,
1998; Smith et al. 1997; Muno et al. 2000; Narita et
al. 2001). Recently, the source became undetectable during the
monitoring observations with the All Sky Monitor (ASM) aboard {\it
RXTE}. In Figure~\ref{fig:lc}, the ASM light curve of KS 1731--260 is
presented, which shows that in early 2001, after a short period during
which solar constraints inhibited monitoring, the source could not be
detected anymore near day 1800. Its disappearance was confirmed by its
non-detection using pointed {\it RXTE} observations with the
proportional counter array (PCA) and the bulge scan observations of
the Galactic center region (Fig.~\ref{fig:lc}; Markwardt 2000;
Markwardt et al. 2000), which include KS 1731--260. The first
non-detection in the bulge scans was on 2001 Feb. 7.  The 1 $\sigma$
upper limit is $1.7 \times 10^{-11}$ \funit~(2--10 keV; absorbed).

Before it disappeared, KS 1731--260 had been accreting for about 12.5
years. This is the only outburst observed for this source and previous
outbursts could have been significantly shorter (less than a year) or
considerably longer (several decades). However, if the behavior
displayed by KS 1731--260 during its last outburst is typical, then
its average outburst duration is of the order of a decade (we use
$t_{\rm o} = 12.5$ years in the following).  To estimate its quiescent
flux using the Brown et al. (1998) model, its time averaged outburst
flux needs to be known.  During all outburst observations, the
absorbed source fluxes were in the range of $10^{-9}$--$10^{-8}$
\funit.  The average absorbed flux obtained during the bulge scans
from 05 Feb 1999 to 30 Oct 2000 was $\sim1.7\times10^{-9}$
\funit~(2--10 keV; with a range of 0.1--4.8 $\times10^{-9}$
\funit). Due to the limited energy ranges for which those fluxes are
quoted and the high column density towards the source, the total
unabsorbed fluxes are likely to be in the range of 0.5--1 $\times
10^{-8}$ \funit, or even higher.  To obtain a lower limit on the
average quiescent time of KS 1731--260, we will assume that its time
averaged outburst flux did not exceed $\sim 10^{-9}$ \funit. It is
conceivable that during previous outbursts the source has reached
similar flux levels (recurrent transients tend to have similar peak
fluxes), so we take as $\langle F_{\rm o} \rangle = 10^{-9}$
\funit. We use this conservative flux to demonstrate later on that
even for such a low flux level, the quiescent time of KS 1731--260
might be several hundreds of years.  Using the Brown et al. (1998)
model, this results in predicted quiescent fluxes of 7.1, 4.3, and
0.85 $\times 10^{-12}$ \funit, for a $t_{\rm q}$ of 1, 10, and 100
years, respectively (see Chen, Shrader, \& Livio 1997 for typical
observed quiescent times and Lasota 2001 for theoretical expected
ones).

\section{Observation, analysis, and results}

We obtained a {\it Chandra}/ACIS-S observation on KS 1731--260 on 27
March 2001 00:17--06:23 UTC (only a few months after it turned off;
Fig.~\ref{fig:lc}) for a total on-source time of $\sim$20 ksec. During
our observation the ACIS-S3 backside-illuminated CCD was used with a
1/4 sub-array. This configuration was chosen to reduce the pile-up
problems which would arise if the source were to exceed a luminosity
of $10^{34}$ \Lunit~($\sim 10^{-12}$ \funit).  No episodes of high
background occurred, so all the data were used. We used the CIAO tools
(version 2.1.3) and the threads listed at http://asc.harvard.edu to
analyze the data. Two point sources were detected using the tool {\it
celldetect}; one of them is located near the center of the {\it
ROSAT}/HRI error circle of KS 1731--260 (Barret et al. 1998) and can
almost certainly be identified with KS 1731--260. For a discussion of
its X-ray position (including the identification of the extra source
with a star in the 2MASS catalog) and its likely optical/infra-red
counterpart, we refer to Wijnands et al. (2001b) and Groot et
al. (2001).

The source spectrum was extracted using a circle of 10 pixels in
radius on the source position. The background data were obtained by
using four background regions, each consisting of a circle with a
radius of 10 pixels. The data were rebinned using the FTOOLS routine
{\it grppha} into bins with a minimum of 10 counts per bin.  The
spectrum was fitted using XSPEC version 11 (Arnaud 1996). We used
several models to fit the data and either the column density, $N_{\rm
H}$, was fixed to the value obtained by {\it ROSAT} and {\it ASCA}
($1.1 \times 10^{22}$ cm$^{-2}$; Barret et al. 1998; Narita et
al. 2001) or it was included in the fit as a free parameter. A pure
power-law and a black-body model fit the data equally well, although a
power-law model gave a very high photon index of 4 ($N_{\rm H}$ fixed)
or 5.3 ($N_{\rm H}$ left free; the resulting value was $\sim 1.7
\times 10^{22}$ cm$^{-2}$). This strongly suggests that the X-ray
spectrum is more likely to be black-body like.  Therefore, we
concentrate on the black-body fits in the rest of our paper.

\begin{figure}
\begin{center}
\begin{tabular}{c}
\psfig{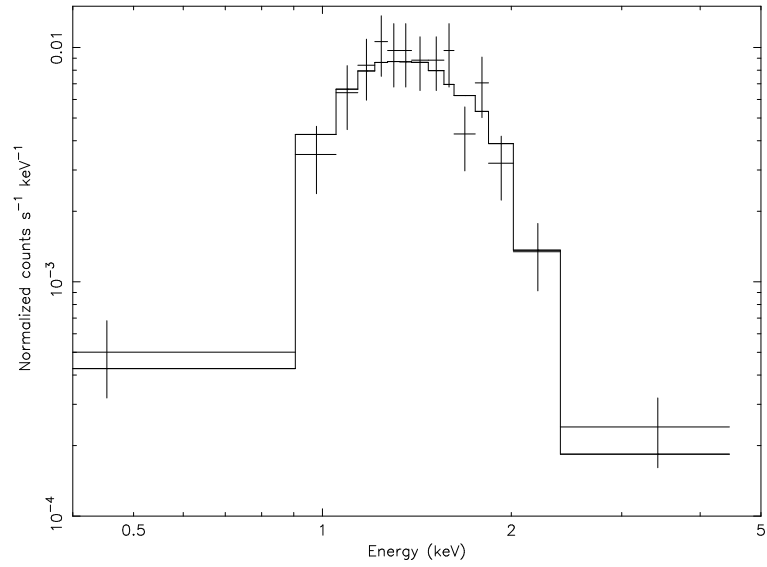}
\end{tabular}
\figcaption{The {\itshape Chandra}/ACIS-S S3 spectrum of KS
1731--260. The solid line is the best black body fit through the data.
\label{fig:spectrum} }
\end{center}
\end{figure}

The obtained spectral results are listed in Table~\ref{tab:spectra}
and the spectrum is shown in Figure~\ref{fig:spectrum}. Because
several quiescent neutron star systems have shown evidence for an
extra power-law component above a few keV in their quiescent spectra
(e.g, Asai et al. 1996, 1998), we also fitted the data with a
black-body plus power-law model (using a fixed photon index of 1;
e.g., Rutledge et al. 2001a). When the column densities are left as a
free parameter, its value is consistent with that obtained from the
{\it ROSAT} and {\it ASCA} data of KS 1731--260. The temperature,
$kT$, of the black body we measure with {\it Chandra}, is in all cases
consistent with 0.3 keV. The unabsorbed 0.5--10 keV black body fluxes
are around $1.7 \times 10^{-13}$ \funit, although the flux is slightly
lower when only a black body model was used with a free floating
column density. The power-law component flux would be $\sim15$\% of
the total flux in the 0.5--10 keV energy range, but it is not
statistically required.

The black body flux results in a 0.5--10 keV luminosity of $\sim 1
\times 10^{33}$ \Lunit~(for 7 kpc; Muno et al. 2000). Rutledge et
al. (2000) found that the bolometric luminosity of quiescent neutron
star systems (using hydrogen atmosphere models) is about a factor of 2
higher than the luminosity in the 0.5--10 keV range (e.g., 4U 1608--52
has a very similar column density and black body temperature in
quiescence [Rutledge et al. 1999] and it has only a factor of two
higher bolometric luminosity compared to the black-body 0.5--10 keV
luminosity; Rutledge et al. 2000). Therefore, we will adopt a
bolometric value of $\sim 2 \times 10^{33}$ \Lunit~in the rest of this
paper, resulting in a quiescent bolometric flux of $\sim 4 \times
10^{-13}$ \funit.

\section{Discussion\label{section:discussion}}

After actively accreting for 12.5 years, the neutron star X-ray binary
KS 1731--260 suddenly turned off in early 2001. It has a quite
different outburst behavior than ordinary transients. In those
ordinary systems, it is believed (e.g., Lasota 2001 and references
therein), that during their quiescent episodes their accretion disks
slowly fill due to low-level mass transfer from the companion star via
Roche-lobe overflow. During outbursts, this matter is dumped onto the
compact object at a rate faster than can be supplied by the donor star
and the disk quickly empties. It will slowly refill again in
quiescence. However, for KS 1731--260 and similar systems, the mass
transfer rate of the companion star can keep up with the mass transfer
rate onto the compact object for a considerable time period (12.5
years for KS 1731--260) and the sources become quasi-persistent with
quasi-stable accretion disks. It is unclear if and how such sources
can be accounted for in the latest versions of the disk instability
models to explain the outbursts in X-ray transients (e.g., Lasota
2001).

We have observed KS 1731--260 in quiescence with {\it Chandra} only a
few months after its transition in order to test the cooling neutron
star model (see Wijnands 2001 for a short discussion about other
models). The bolometric flux of KS 1731--260 in quiescence is $\sim 4
\times 10^{-13}$ \funit, resulting in a bolometric luminosity of $\sim
2 \times 10^{33}$ \Lunit~(for a distance of 7 kpc; Muno et
al. 2000). This bolometric luminosity and the black body temperature
($kT \sim0.3$ keV) are very similar to what has been observed for
other, short-duration transients (see, e.g., Asai et al. 1996, 1998;
Rutledge et al. 2001a, 2001b).  This result is unexpected, because, as
explained in Section~\ref{section:intro}, the long accreting episode
should have heated the neutron star in KS 1731--260 to a higher
temperature (and thus higher luminosity) than what we have
observed. However, this assumes that (i) the Brown et al. (1998) model
for short duration transients also applies to systems such as KS
1731--260, (ii) standard cooling operates in the neutron star in KS
1731--260, and (iii) the source spends a similar amount of time in
quiescence as the other known X-ray transients.

Therefore, one obvious solution to this discrepancy would be that the
time KS 1731--260 spends in quiescence is extremely long compared to
other systems. Our general believe on how long transients can be
quiescent is based on those systems for which multiple outbursts have
been observed and is therefore biased to short quiescent episodes
(years to a few decades, basically the time since the birth of X-ray
astronomy). For certain types of transients, the quiescent episodes
could last significantly longer. However, if the standard cooling
model for neutron stars applies here, KS 1731--260 has to be quiescent
in between outbursts for at least 200 years. The quiescent intervals
most likely have to be considerably longer. To calculate the expected
quiescent flux, we used a very conservative bolometric outburst flux
of $10^{-9}$ \funit, but the true value is more likely in the range of
$0.5-1 \times 10^{-8}$ \funit, or even larger. If indeed this is a
typical outburst flux level, this would indicate that the source has
to spend over a thousand years in quiescence. In the present day disk
instability models (e.g., Lasota 2001; Dubus, Hameury, \& Lasota
2001), it is not easy to see how such long quiescent episodes can
occur because the accretion disk will slowly fill again, but at a
certain point, even small fluctuations in the mass transfer rate from
the companion star will trigger an outburst (see, e.g., Lasota 2001
for a discussion). An alternative model for this source might be the
companion star instability model.

But if some LMXBs can indeed be quiescent for over a thousand years,
then there may be a considerable number of such systems in our
Galaxy. Only a rough number estimate can be obtained by assuming that
such systems are active on average for ten years and quiescent for
thousand years. At the moment, we know of only a few sources which
fall in the class of the long-duration transients and which where
actively accreting during the last decade (e.g., KS 1731--260, MXB
1659--29, EXO 0748--676, GS 1826--238).  Thus, if these assumptions
are correct, one would expect not more than several hundred of such
systems.  Although the numbers are highly uncertain, if indeed such a
large number of extra previously unrecognized LMXBs are present in our
Galaxy, then they might help to bring the birth rate of binary
millisecond radio pulsars and of LMXBs in better agreement (Kulkarni
\& Narayan 1988; Cot\'e \& Pylyser 1989; see Lorimer 1999 for a recent
discussion).

Alternatively, instead of postulating that the quiescent episodes of
KS 1731--260 are extremely long, one could also speculate on the
possibility that enhanced cooling takes place in the core of the
neutron star in this system. If strong enhanced neutrino cooling
occurs in the core (Colpi et al. 2001; Ushomirsky \& Rutledge 2001),
this might cool down the star in a much shorter time span and this
could possibly explain why KS 1731--260 is so underluminous in
quiescence compared to the expected value inferred from cooling
neutron star models (Brown et al. 1998). Colpi et al. (2001) suggested
that such a mechanism is at work in the neutron star transient Cen
X-4, to explain its low ratio of quiescent to outburst
luminosity. They also suggested that this enhanced cooling might be
caused by a more massive neutron star in Cen X-4 compared to those in
the other systems, which would suggest that a massive neutron star
might also be present in KS 1731--260.

Finally, we have assumed that the cooling neutron star model developed
by Brown et al. (1998) for the short duration transients, is valid
also for the long-duration transients such as KS 1731--260. However,
this model might break down for systems like KS 1731--260. It is
unclear what the effects of such a long active period are on the crust
and the core of the neutron star. It is plausible that the crust looks
more like that of a neutron star in a persistent X-ray binary. The
X-ray flux observed from KS 1731--260 might be dominated by the state
of the crust.  However, the core should then be even cooler than we
have assumed, which would strengthen our conclusion that KS
1731--260 has to be in quiescence for a very long time.  More detailed
modeling of the behavior of the neutron star (its crust and its core)
needs to be performed to understand long-duration systems like KS
1731--260.

{\itshape Note added in manuscript} We became aware of the work of
Burderi et al. (2001) and Rutledge et al. (2001c), using {\it
BeppoSAX} and our {\it Chandra} quiescent observations of KS
1731--260, respectively. Both papers confirm our conclusion: if the
X-ray emission originates from the neutron star surface, KS 1731--260
has to have been in quiescence for $>$1000 years.

\clearpage

\clearpage
\clearpage

\begin{deluxetable}{lcc}
\tablecolumns{3}
\tablewidth{0pt} 
\tablecaption{Spectral results$^a$\label{tab:spectra}}
\tablehead{\multicolumn{3}{c}{Black body}}
\startdata
$N_{\rm H}$ ($10^{22}$ cm$^{-2}$) & $0.9^{+0.4}_{-0.2} $   & 1.1 (fixed) \\
$kT$ (keV)                        & $0.30^{+0.05}_{-0.06}$ & 0.27\pp0.03 \\
Norm. ($10^{-6}$)                 & $1.7^{+2.1}_{-1.7}$    & $2.5^{+0.7}_{-0.5}$ \\
$F$ ($10^{-13}$ \funit)           & $1.2$                  & $1.7 $ \\
$\chi^2$/dof                      & 8.0/13                 & 9.0/14 \\
\hline
\multicolumn{3}{c}{Black body + power law$^b$}\\
\hline
$N_{\rm H}$ ($10^{22}$ cm$^{-2}$) & 1.1\pp0.4             & 1.1 (fixed) \\
$kT$ (keV)                        & 0.25\pp0.06           & 0.25\pp0.03 \\
Norm. black body ($10^{-6}$)      & $2.8^{+8.4}_{-1.7}$   & $2.7^{+0.6}_{-0.5}$ \\
Norm. power law ($10^{-6}$)       & $2.3^{+1.5}_{-2.0}$   & $1.9^{+1.4}_{1.5}$\\
$F_{\rm tot}$ ($10^{-13}$ \funit) & $2.0$                 & $2.1 $ \\
$F_{\rm bb}$ ($10^{-13}$ \funit)  & $1.7$                 & $1.8 $ \\
$F_{\rm pl}$ ($10^{-13}$ \funit)  & $0.3$                 & $0.3 $ \\
$\chi^2$/dof                      & 5.6/12                & 4.9/13 \\
\enddata 
\tablenotetext{a}{The error bars are for 90\% confidence levels. The
fluxes are unabsorbed and for 0.5--10 keV}
\tablenotetext{b}{The power law photon index was fixed to 1}
\end{deluxetable}

\end{document}